\documentclass[12pt,a4paper]{article}
\usepackage{jheppub}
\usepackage{amsmath,amssymb,amsfonts,mathtools,amsmath,amssymb,amsthm,amscd} 
\usepackage{float}
\usepackage{array}
\usepackage{wrapfig}
\usepackage{fancybox}
\newcommand{\bea}{\begin{eqnarray}\displaystyle}
\newcommand{\eea}{\end{eqnarray}}

\newlength{\arrow}
{\setlength{\fboxsep}{15pt}
\setlength{\mylength}{\linewidth}%
\addtolength{\mylength}{-2\fboxsep}%
\addtolength{\mylength}{-2\fboxrule}%
\Sbox
\minipage{\mylength}%
\setlength{\abovedisplayskip}{0pt}%
\setlength{\belowdisplayskip}{0pt}%
\equation}%
{\endequation\endminipage\endSbox
\[\fbox{\TheSbox}\]}

\newdimen\tableauside\tableauside=1ex
\newdimen\tableaurule\tableaurule=0.8pt
\newdimen\tableaustep
\def\phantomhrule#1{\hbox{\vbox to0pt{\hrule height\tableaurule width#1\vss}}}
\def\phantomvrule#1{\vbox{\hbox to0pt{\vrule width\tableaurule height#1\hss}}}
\def\sqr{\vbox{%
  \phantomhrule\tableaustep
  \hbox{\phantomvrule\tableaustep\kern\tableaustep\phantomvrule\tableaustep}%
  \hbox{\vbox{\phantomhrule\tableauside}\kern-\tableaurule}}}
\def\squares#1{\hbox{\count0=#1\noindent\loop\sqr
  \advance\count0 by-1 \ifnum\count0>0\repeat}}
\def\tableau#1{\vcenter{\offinterlineskip
  \tableaustep=\tableauside\advance\tableaustep by-\tableaurule
  \kern\normallineskip\hbox
    {\kern\normallineskip\vbox
      {\gettableau#1 0 }%
     \kern\normallineskip\kern\tableaurule}%
  \kern\normallineskip\kern\tableaurule}}
\def\gettableau#1{\ifnum#1=0\let\next=\null\else
\squares{#1}\let\next=\gettableau\fi\next}

\begin{document}

\hfill ITEP-TH-08/18

\hfill IITP-TH-06/18

\title{Refined Topological Branes}
\author[a,b,c,1]{Can Koz\c{c}az\note{Current affiliation Bo\u{g}azi\c{c}i University}}
\author[d,e,f,2]{Shamil Shakirov\note{Current affiliation Mathematical Sciences Research Institute}}
\author[c]{Cumrun Vafa}
\author[g,b,c,3]{and Wenbin Yan\note{Current affiliation Yau Mathematical Sciences Center}}
\affiliation[a]{Department of Physics, Bo\u{g}azi\c{c}i University\\
34342 Bebek, Istanbul, Turkey}
\affiliation[b]{Center of Mathematical Sciences and Applications, Harvard University\\
20 Garden Street, Cambridge, MA 02138, USA}
\affiliation[c]{Jefferson Physical Laboratory, Harvard University\\
17 Oxford Street, Cambridge, MA 02138, USA}
\affiliation[d]{Society of Fellows, Harvard University\\
Cambridge, MA 02138, USA}
\affiliation[e]{Institute for Information Transmission Problems, Moscow 127994, Russia}
\affiliation[f]{Mathematical Sciences Research Institute, Berkeley, CA 94720, USA}
\affiliation[g]{Yau Mathematical Sciences Center,\\
Tsinghua University, Haidian district, \\
Beijing, China, 100084}

\abstract{We study the open refined topological string amplitudes using the refined topological vertex. We determine the refinement of holonomies necessary to describe the boundary conditions of open amplitudes (which in particular satisfy the required integrality properties). We also derive the refined holonomies using the refined Chern-Simons theory. }

\maketitle
\section{Introduction}

\par{Gauge theories with ${\cal N}=2$ supersymmetry in 4d have been important playground for theoretical physics since the celebrated solution of Seiberg and Witten \cite{Seiberg:1994rs,Seiberg:1994aj}. They can be geometrically engineered in type IIA string theory \cite{Katz:1996fh,Katz:1996th}. For theories with $SU(N)$ gauge group,  compactification manifolds are known to be local toric Calabi-Yau threefolds, and the closed topological string theory on these threefolds encodes essential information about the low energy dynamics of gauge theories. The genus zero topological string amplitude on them determines the prepotential of gauge theories. Higher genus amplitudes further encode gravitational couplings and to a large extent are captured by their holomorphic structure \cite{Bershadsky:1993cx}. The topological vertex \cite{Aganagic:2003db}  (see also \cite{Iqbal:2002we}) and its refinement \cite{Awata:2005fa,Iqbal:2007ii} solve the problem of computing all genus string amplitudes on local toric threefolds.}

\par{Topological string theory has been shown to compute contributions coming from inserting codimension two surface operators with certain singularities of the gauge bundle \cite{Kozcaz:2010af,Dimofte:2010tz}. They are realized by wrapping D4 branes on Lagrangian submanifolds inside the internal space, and extending them along ${\mathbb R}^{2}\subset {\mathbb R}^{4}$. For theories with $SU(N)$ gauge group,  we can place the topological branes either on the external or internal legs of the associated toric diagram describing the compactification manifold. The branes provide boundary conditions for holomorphic curves from punctured worldsheet into the target space, and the open topological strings count such maps. The topological vertex in the unrefined case also solves the problem of computing all genus amplitudes for external and internal branes. However, the open amplitude computation turns out to be more subtle when we consider the refined topological string theory, especially when we want to focus on internal branes.}

\par{A widely adopted approach in formulating refined open topological amplitudes is based on refining holonomies induced on branes, in addition to the more recent approaches using geometric transition \cite{Gopakumar:1998ki} in the refined context \cite{Mori:2016qof,Kimura:2017auj,Kameyama:2017ryw}. As it is very common in the refinement, holonomies which are given in terms of Schur functions are replaced by their refined analog, Macdonald functions. This approach looks very plausible and natural, but once the free energy is studied more thoroughly one encounters inconsistencies. This can be observed even in the simplest possible brane configuration: a single internal brane (i.e. one D4 brane) on the resolved conifold. The brane is  wrapped on a Lagrangian submanifold on the resolved conifold, and we need to make a choice whether the remaning dimensions of the D4 brane extends either on the plane acted by equivariant parameter $q=e^{i\varepsilon_{1}}$ or $t=e^{-i\varepsilon_{2}}$. We call it a $q$- or $t$-brane depending which ${\mathbb R}^{2}$ it occupies.  After our choice, the brane can not lie on both planes at the same time. Therefore, we expect the amplitude to depend on one of the two equivariant parameters, in addition to the holonomy induced on the brane. However, the existing approach based on naive replacement of the holonomies described above results in an amplitude that depends on both equivariant parameters, a clear indication that the formalism should be modified.}

\par{Another important clue can be obtained by placing a stack of branes along one of the refined topological vertex $C_{\lambda\mu\nu}(t,q)$, say the first leg $\lambda$. If we choose the holonomy to be expressed in terms of the Schur function, then it is a $t$-brane. Nevertheless, there is no reason why we should not be able to wrap D4 branes on this Lagrangian submanifold and extend them on the other ${\mathbb R}^{2}$ plane, making it a $q$-brane. We will address these questions in this short note, and demonstrate that physically consistent results can be obtained by modifying holonomies from Schur functions not only to Macdonald functions but also to dual Macdonald functions and dual Schur functions with respect to the $(q,t)$-inner product defined by Macdonald \cite{macdonald}. We adopt two different approaches to derive the correct holonomies\footnote{There is an alternative but equivalent approach to obtain the correct refined open amplitudes. We could keep the holonomies unrefined, but change instead the refined propagator. In this note, we take the former approach. }: one, which relies on the integral expansion of the open topological string free energy, and another based on refined Chern-Simons theory.  All refined holonomies we formulate reduce to Schur functions in the unrefined limit as they should.   }

\par{As a byproduct, we apply our formalism to reproduce the explicit form of dual elliptic Macdonald functions from toric geometries. In \cite{Hollowood:2003cv}, partial and full compactifications of toric webs are introduced, and it is argued that they correspond to elliptic and dual fibrations respectively. We call them horizontal and vertical compactifications, and associate them with position and momentum variables respectively, borrowing terminology from integrable systems. The variable associated with the compactification is elliptically deformed. In this paper, we only focus on placing branes along the horizontal legs. We study both the horizontal compactification in which the brane is an internal brane and the position variable becomes elliptic. In addition, we focus on the vertical compactification when the brane becomes an external brane and the momentum variable has elliptic dependence.     }

\par{In section \ref{sec:topologicalBranes}, we review the open topological string computation for the unrefined case when the brane is placed on an internal leg of the toric diagram. Then we propose how this computation should be refined, and show explicitly that our proposal produces desired form for the free energies. In section \ref{sec:rCS}, we reproduce the refined holonomies using the annuli amplitudes of the refined Chern-Simons theory. In section \ref{sec:elliptic}, we use our formalism to reproduce the dual elliptic Macdonald functions.  In section \ref{sec:discussion}, we summarize and discuss our results. In Appendix, we collect some relevant information about symmetric functions. }



\section{Topological Branes}
\label{sec:topologicalBranes}

Computing open topological string partition functions on Calabi-Yau threefolds in the presence of D4 branes is an important problem from several perspectives. Geometrically, these topological string amplitudes encode invariants of holomorphic curves with boundaries inside the Calabi-Yau threefold. In terms of type IIA string theory/M-theory, they are counting BPS degeneracies of D2/M2 branes winding holomorphic cycles with boundaries on D4/M5 branes which wrap Lagrangian cycles of the threefold and extend in ${\mathbb R}^{2}\subset{\mathbb R}^{4}$. From the perspective of gauge theory living on ${\mathbb R}^{4}$, they are partition functions of codimension two surface operators.

Historically, this problem has been first addressed and solved in the context of unrefined topological strings. The amplitudes depend on boundary conditions for holomorphic maps, which can be expressed  either by conjugacy classes of symmetric group (reflecting how the discs ending on D4-branes are wrapping the $S^{1}$) or equivalently in terms of representations of $U(\infty)$ labelled by arbitrary Young diagrams of any number of rows and columns. On D4 branes, the maps induce a holonomy $V$ for the gauge connection supported by branes. The trace of $V$ in a representation $\mu$ of the unitary group is given by the Schur function of eigenvalues $(v_1, \ldots, v_n)$ of $V$,
\begin{align}
\mbox{tr}_{\mu}V=s_{\mu}(V)=s_{\mu}(v_1, \ldots, v_n).
\end{align}

Two distinct possibilities arise for the location of the D4 branes on a local toric Calabi-Yau threefold: either, they can be placed either along  external or internal legs of the associated toric diagram. Accordingly, one talks about external or internal branes. In both cases, a simple prescription is available for the partition function using the topological vertex. The contribution of an edge with an external brane is given by
\begin{align}
\begin{array}{ccc}\includegraphics[width=0.1\textwidth]{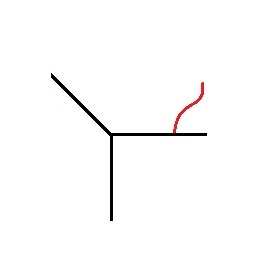}\end{array} :\hspace{10ex}Z^{open}(V)= \sum\limits_{\mu} \ C_{\lambda \sigma\mu^t}(q) \ f_{\mu}(q)^p \ {\mbox tr}_{\mu}\,V,
\label{ext}
\end{align}
where $f_{\mu} (q)= (-1)^{|\mu|} q^{\Arrowvert\mu\Arrowvert^2/2 - \Arrowvert\mu^t\Arrowvert^2/2}$ is the framing factor and $p$ is the amount of framing of the brane. On the other hand, the contribution due to the internal brane is more involved and given by
\begin{align}\nonumber
\begin{array}{ccc}\includegraphics[width=0.2\textwidth]{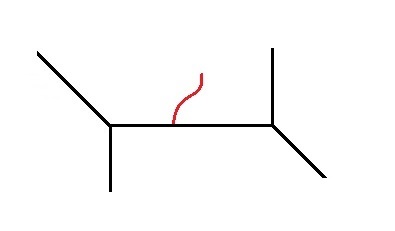}\end{array} :\hspace{3ex}Z^{open}(V)&= \sum_{\mu,\mu_{L},\mu_{R}}(-Q)^{|\mu|}Q_{L}^{|\mu_{L}|}Q_{R}^{|\mu_{R}|}\,C_{\lambda \sigma  (\mu\otimes\mu_{L})}(q)C_{\tau \eta (\mu^{t}\otimes\mu_{R})}(q)\\ &\times\mbox{tr}_{\mu_{L}}V\,\mbox{tr}_{\mu_{R}}V^{-1},
\label{int}
\end{align}
where $Q_L$ and $Q_R$ denote the K\"{a}hler parameters associated to the sizes of the left and right discs with respect to the internal brane, respectively, and $Q=Q_{L}Q_{R}$ is K\"{a}hler parameter of the whole ${\mathbb P}^{1}$. $\mbox{tr}_{\mu_{L}}V$ is the holonomy induced on the brane from the disc left ending on it, and it describes the boundary conditions for the open amplitudes, and likewise $V^{-1}$ is due to the disc from the right of the brane\footnote{The difference of exponents is due to opposite windings of holomorphic maps.}.

Above expressions are useful, as they allow us to study various geometric and gauge-theoretic properties of interest. For example, in the presence of external branes the open topological string partition function is conjectured to possess the following integral structure \cite{Ooguri:1999bv,Labastida:2000yw},
\begin{align}
Z(V)=\exp\left( \sum_{k=1}^{\infty}\sum_{s,\beta,\mu}\frac{1}{k}N_{\mu,\beta,s}\frac{1}{q^{k/2}-q^{-k/2}}Q_{\beta}^{k}q^{ks}\,\mbox{tr}_{\mu}V^{k}\right),
\end{align}
where $N_{\mu,\beta,s}$'s are a priori undetermined integers and are the degeneracies of BPS states due to D2 branes wrapping a relative homology class $\beta\in H_{2}(X,{\mathcal L})$ with the Lagrangian submanifold ${\mathcal L}$. They have the 2d space-time spin $s$ and $SU(M)$ representation $\mu$. One can compute these integers up to any desired order once the open topological string amplitude is computed by any means.

As alluded in the Introduction, the situation is more subtle in the refined case. For unrefined topological string theory, the partition function is insensitive to the choice of ${\mathbb R}^{2}\subset{\mathbb R}^{4}$ that the D4 branes occupy. However, if the space-time is subject to $\Omega$-deformation, the planes are distinguished by different rotations. $\Omega$-background introduces a complex structure to ${\mathbb R}^{4}\simeq {\mathbb C}^{2}$ and rotates different coordinates by $\varepsilon_{1}$ and $\varepsilon_{2}$:
\begin{align}\label{omega}
z_{1}\mapsto e^{i\varepsilon_{1}}z_{1}\coloneqq\,q\,z_{1} ,\qquad z_{2}\mapsto e^{i\varepsilon_{2}}z_{2}\coloneqq\,t^{-1}\,z_{2}.
\end{align}
The refined topological string distinguishes between the two ${\mathbb R}^{2}$ planes. Depending on which plane the D4 brane is extending, we talk about $q$- or $t$-branes. Such a distinction is not present in the unrefined topological string. The corresponding open topological string free energy has an integral expansion in terms of $N_{\mu,\beta,s_{L},s_{R}}$, which reads as
\begin{align}
Z=\exp\left(\sum_{k=1}^{\infty}\sum_{s_{L},s_{R},\beta,\mu}\frac{1}{k}N_{\mu,\beta,s_{L},s_{R}}\frac{1}{x^{k/2}-x^{-k/2}}q^{ks_{L}}t^{-ks_{R}} Q_{\beta}^{k}\,\mbox{tr}_{\mu}V^{k}\right),
\end{align}
where $x$ is either $q$ or $t$ depending on whether we insert a $q$- or $t$-brane. The extra grading can be best understood when we lift to M-theory, as it is the case for closed amplitudes \cite{Gopakumar:1998ii,Gopakumar:1998jq}. The BPS particles are charged under the little group of massive particles $SO(4)\simeq SU(2)_{L}\times SU(2)_{R}$. We measure  left $s_{L}$ and right $s_{R}$ spins by $q$ and $t$ parameters, respectively.

\subsection{External topological branes}
\label{sec:ExtTopBranes}

\paragraph{}We start our discussion with a single stack of external branes on different legs of the refined topological vertex. It is desirable to have a refined version of equations Eq. \ref{ext} and Eq. \ref{int} to compute refined integer invariants. Such a computation, in addition to the subtlety arising from $q$ versus $t$-branes, would be also sensitive to whether the brane is located along the preferred or un-preferred leg of the refined topological vertex. We will now demonstrate that the choice of holonomy polynomials $\mbox{tr}_{\mu}\,V$ should be ``refined'' \cite{Kozcaz:2010af, Dimofte:2010tz, Iqbal:2011kq}. The need for such a change can easily be seen along the preferred direction: assuming that the holonomy is still given by Schur functions, the integrality of the open free energy is lost:
\begin{align}\label{wrong}\nonumber
&F=\log Z(V)=\log \left( \sum_{\nu}\Lambda^{|\nu|}C_{\emptyset\emptyset\nu}(t,q)(-1)^{|\nu|}s_{\nu}(V)\right)\\\nonumber
&=\log \left( \sum_{\nu}\Lambda^{|\nu|}(-
v)^{|\nu|}P_{\nu}(t^{\rho};q,t)(-1)^{|\nu|}s_{\nu}(V)\right)\\
&= \Lambda\, \frac{q^{1/2}t^{-1/2}}{t^{1/2}-t^{-1/2}}\mbox{tr}_{\tableau{1}}V+\Lambda^{2}\frac{qt^{-1} }{2(t-t^{-1})}\mbox{tr}_{\tableau{1}}V^{2}+\Lambda^{2}\frac{q(q-t)}{2(1-t)(1-t^{2}) (1 - q t)}\mbox{tr}_{\tableau{2}}V+\mathellipsis,
\end{align}
where problematic terms deviating from the generic structure clearly vanish when $t=q$. $\Lambda$ denotes the size of the disc, which could have been absorbed too by rescaling $V$; we chose not to. This pathology can be cured if we change the holonomy to be given by the Macdonald polynomial $v^{-|\nu|}\imath P_{\nu^{t}}(V;t,q)$,
\begin{align}\nonumber
Z(V)&=\sum_{\nu}\Lambda^{|\nu|}(-v)^{|\nu|}P_{\nu}(t^{\rho};q,t)v^{-|\nu|}\,\imath P_{\nu^{t}}(V;t,q)\\
&=\prod_{i,j}\left (1-\Lambda\,t^{\rho_{i}}V_{j} \right)^{-1}=\exp\left(\sum_{d=1}^{\infty}\frac{1}{d} \frac{\Lambda^{d}}{t^{d/2}-t^{-d/2}}\mbox{tr}_{\tableau{1}}V^{k}\right).
\end{align}

\par{The denominator in the exponential indicates that the deformed holomony gives rise to $q$-branes. However, nothing prevents us equally wrapping D4 branes on the Lagrangian submanifold along the preferred direction and extend them on the ${\bar t}$-plane. Therefore, the natural question, as it was mentioned in the introduction, is how to obtain the amplitude for a stack of $t$-branes. The refined topological vertex has a given form, hence, the only freedom we have at our disposal is to change the holonomy. Macdonald defined so-called dual Macdonald polynomials $Q_{\mu}(V;t,q)$ where the duality is defined with respect to the $\langle\cdot,\cdot\rangle_{q,t}$ inner product.  If we take the holonomy to be proportional to dual Macdonald function $Q_{\mu^{t}}$, we have }
\begin{align}\nonumber
Z(V)&=\sum_{\nu}\Lambda^{|\nu|}(-v)^{|\nu|}P_{\nu}(t^{\rho};q,t)(-1)^{|\nu|} Q_{\nu}(V;q,t)\\
&=\prod_{i,j}\left (1-\Lambda\,q^{\rho_{i}}V_{j} \right)^{-1}=\exp\left(\sum_{d=1}^{\infty} \frac{1}{d}\frac{\Lambda^{d}}{q^{d/2}-q^{-d/2}}\mbox{tr}_{\tableau{1}}V^{k}\right).
\end{align}
Note that the choice of the holonomy also depends on the way we label the refined topological vertex, but once a choice is made then the holonomy is fixed by the integrality condition.

Similarly, we can place a stack of either $q$- or $t$-branes along one of the two un-preferred directions. So far in the literature the holonomies along these legs are only taken to be Schur funtions, which are independent of $q$ and $t$. Up to framing factors, the refined topological vertex depends on one of equivariant parameters if we wrap branes on one of the un-preferred directions. Augmenting with Schur functions as holonomies suggest that along one of un-preferred directions we can only have $q$-branes and along the other one only $t$-branes. Again, this picture does not capture the whole physics and therefore is too restrictive. Equipped with our new understanding on the choice of holonomies along the preferred direction, let us determine the holonomies for un-preferred direction. Imagine we want to compute the open amplitude after placing a stack of branes along the first leg of the vertex, $C_{\lambda\emptyset\emptyset}(t,q)$:
\begin{align}\nonumber
Z(V)&=\sum_{\lambda}\Lambda^{|\lambda|}v^{|\lambda|}s_{\lambda^{t}}(t^{-\rho})(-v)^{-|\lambda|}s_{\lambda}(V)\\
&=\prod_{i,j}(1-\Lambda\,t^{-\rho_{i}}V_{j})=\exp\left(\sum_{d=1}^{\infty}\frac{1}{d} \frac{\Lambda^{d}}{t^{d/2}-t^{-d/2}}\mbox{tr}_{\tableau{1}}V^{k}\right),
\end{align}
which is the partition function for a $t$-brane. We can get a $q$-brane if we chose the holonomy to be proportional to the dual of the Schur function with respect to the $(q,t)$-inner product; $S_{\lambda}(V;q,t)$\footnote{The definition of the dual Schur function $S_{\lambda}(V;q,t)$ is given in Eq. \ref{dualschur} in the Appendix.}:

\begin{align}\nonumber
Z(V)&=\sum_{\lambda}\Lambda^{|\lambda|}v^{|\lambda|}s_{\lambda^{t}}(t^{-\rho})(-1)^{|\lambda|}S_{\lambda}(V;q,t)\\
&=\prod_{i,j}(1-\Lambda\,q^{-\rho_{i}}V_{j})=\exp\left(\sum_{d=1}^{\infty}\frac{1}{d} \frac{\Lambda^{d}}{q^{d/2}-q^{-d/2}}\mbox{tr}_{\tableau{1}}V^{k}\right),
\end{align}
which is clearly a $q$-brane partition function. As we will elaborate later in subsection \ref{dualitysection}, there is a nice relationship between the holonomies  $v^{-|\nu|}\imath P_{\nu^{t}}(V;t,q)$ and $(-1)^{|\nu|} Q_{\nu}(V;q,t)$ along the preferred direction, and the same relationship is true for holonomies $(-v)^{-|\lambda|}s_{\lambda}(V)$ and $(-1)^{|\lambda|}S_{\lambda}(V;q,t)$ along the un-preferred direction.

\subsection{Internal topological branes}
\label{sec:IntTopoBra}

Let us turn our attention to internal branes, i.e. topological branes placed along internal legs of the toric diagram. For the usual topological string theory, a local prescription is given for the computation of their contributions \cite{Aganagic:2003db}: one can isolate the internal leg carrying the brane, and compute the associated disc amplitudes locally using,
\begin{align}\label{unrefinedinternal}
\sum_{\mu,\mu_{L},\mu_{R}}(-Q)^{|\mu|}Q_{L}^{|\mu_{L}|}Q_{R}^{|\mu_{R}|}\,C_{\lambda\sigma(\mu\otimes\mu_{L})}(q)C_{\tau\eta(\mu^{t}\otimes\mu_{R})}(q)\,\mbox{tr}_{\mu_{L}}V\,\mbox{tr}_{\mu_{R}}V^{-1},
\end{align}
\smallskip\\
where $Q=Q_{L}Q_{R}$, and $V$ is the holonomy on the brane from disc ending on it from left, and likewise $V^{-1}$ is due to  discs from the right of the brane. In the above expression, for simplicity we ignore possible framing factors due to line bundles over ${\mathbb P}^{1}$ which are different than ${\cal O}(-1)\oplus{\cal O}(-1)\mapsto {\mathbb P}^{1}$, and due to the relative orientation of the interanl brane in the geometry. The representations $\lambda$, $\sigma$, $\tau$ and $\eta$ allow us to glue this leg with the internal brane to the rest of the geometry.

Although the computation of the open refined topological string partition function includes subtleties related to holonomies, the partition function is expected to algebraically obey similar gluing rules. However, as we pointed out before, we need to choose what type of branes we study, and refine holonomies accordingly.  We will seperate the study of $q$ and $t$-branes in  following subsections, and propose refined holonomies based on the requirement that the open free energy has an integral expansion in the refined case as well. Later we will reproduce  same holonomies using refined Chern-Simons theory and geometric transition in section \ref{sec:rCS}.

\subsubsection{Topological $t$-branes}
\label{sec:Topot-branes}

In this subsection, we refine the internal brane computation while keeping the algebraic structure the same as in Eq. \ref{unrefinedinternal}. First, we demonstrate that the natural generalization of holonomies from Schur polynomials to Macdonald polynomials leads to a contradiction. Later, we use the integrality requirement for the free energy  and predict the correct refinement of holonomies.

Let us first keep  holonomies unknown and perform the refined computation explicitly as much as we can. We will assume that the internal brane is placed along the preferred direction of the refined topological vertex,
{\fontsize{10pt}{0pt}
\begin{align}\label{refinedt}\nonumber
Z(V)&=\sum_{\mu,\mu_{L},\mu_{R}}(-Q)^{|\mu|}Q_{L}^{|\mu_{L}|}Q_{R}^{|\mu_{R}|}C_{\emptyset\emptyset (\mu\otimes \mu_{L})}(t,q)C_{\emptyset\emptyset (\mu^{t}\otimes \mu_{R})}(q,t)\,\mbox{tr}_{\mu_{1}}V\, \mbox{tr}_{\mu_{2}}V^{-1}\\\nonumber
&=\sum_{\mu,\mu_{L},\mu_{R}}(-Q)^{|\mu|}Q_{L}^{|\mu_{L}|}Q_{R}^{|\mu_{R}|}(-v)^{|\mu\otimes \mu_{L}|}P_{\mu\otimes \mu_{L}}(t^{\rho};q,t)(-v)^{-|\mu\otimes \mu_{R}|}P_{\mu^{t}\otimes \mu_{R}}(q^{\rho};t,q)\\\nonumber
&\times\mbox{tr}_{\mu_{L}}V\, \mbox{tr}_{\mu_{R}}V^{-1}\\\nonumber
&=\sum_{\mu}(-Q)^{|\mu|} \left(\sum_{\mu_{L},\eta_{L}} (-vQ_{L})^{|\mu_{L}|} {\widehat N}^{\eta_{L}}_{\mu\mu_{L}}(q,t) P_{\eta_{L}}(t^{\rho};q,t)\,\mbox{tr}_{\mu_{L}}V \right)\\
&\qquad\qquad\qquad\,\, \times \left(\sum_{\mu_{R},\eta_{R}} (-v^{-1}Q_{R})^{|\mu_{R}|} {\widehat N}^{\eta_{R}}_{\mu^{t}\mu_{R}}(t,q) P_{\eta_{R}}(q^{\rho};t,q)\,\mbox{tr}_{\mu_{R}}V^{-1} \right).
\end{align}
The refined topological vertex with trivial un-preferred and non-trivial preferred direction is proportional to the Macdonald polynomial. Therefore, we use the refined Littlewood-Richardson coefficents $ {\widehat N}$ for the tensor product, which are rational functions of $q$ and $t$ and reduce to the usual Littlewood-Richardson coefficients when we take the unrefined limit $q=t$.
\par{Just as in the external brane example, the crucial point here is the choice of holonomy polynomials $\mbox{tr}_{\mu_{L}}V$ and $\mbox{tr}_{\mu_{R}}V^{-1}$. Equipped with the previous discussion about the $q$- and $t$-branes, we pick  correct holonomies. The sums in parentheses are computing the disc amplitudes on each side of the internal branes. If we are studying an internal $t$-brane, we can pick}
\begin{align}
\boxed{\mbox{tr}_{\mu_{L}}V =v^{-|\mu_{L}|}\imath P_{\mu_{L}^{t}}(V;t,q),}
\end{align}
and we can perform the $\mu_{L}$-sum first,
\begin{align}
\sum_{\mu_{L}} (-vQ_{L})^{|\mu_{L}|} {\widehat N}^{\eta_{L}}_{\mu\mu_{L}}(q,t)\, v^{-|\mu_{L}|}\imath P_{\mu_{L}^{t}}(V;t,q)=\imath P_{\eta_{L}^{t}/\mu^{t}}(-Q_{L}\,V;t,q),
\end{align}
and then we peform the remaining $\eta_{L}$-sum leading to
\begin{align}\label{leftdisc}
 \sum_{\eta_{L}}P_{\eta_{L}}(t^{\rho};q,t)\imath P_{\eta_{L}^{t}/\mu^{t}}(-Q_{L}\,V;t,q)=P_{\mu}(t^{\rho};q,t)\exp\left (\sum_{d=1}^{\infty}\frac{1}{d}\frac{Q_{L}^{d}}{t^{d/2}-t^{-d/2}}\mbox{tr}_{\tableau{1}}V^{d} \right ),
\end{align}
which has the correct contribution to the free energy from the left disc, consistent with the results in \cite{Ooguri:1999bv}.

Next we need to determine how $\mbox{tr}_{\mu_{R}}V^{-1}$ should be refined for a $\bar{t}$-brane insertion. One might expect that the holomony is the same as the one for the left disc, possibly up to a $q\leftrightarrow t$ exchange. However, this choice for the right holonomy fails to produce the correct free energy. Similar to the computation of the left dics amplitude,  we  first perform the $\mu_{R}$-sum for the right disc in Eq. \ref{refinedt},
\begin{align}
\sum_{\mu_{R}} (-v^{-1}Q_{R})^{|\mu_{R}|} {\widehat N}^{\eta_{R}}_{\mu^{t}\mu_{R}}(t,q)v^{|\mu_{R}|} \imath P_{\mu_{R}^{t}}(V^{-1};q,t)=\imath P_{\eta_{R}^{t}/\mu}(-Q_{R}V^{-1};q,t),
\end{align}
and then the $\eta_{R}$-sum. The contribution from the right disc can again be written as a plethystic exponential,
\begin{align}\nonumber
\sum_{\eta_{R}}P_{\eta_{R}}(q^{\rho};t,q)\imath P_{\eta_{R}^{t}/\mu}(-&Q_{R}V^{-1};q,t)\\
&=P_{\mu^{t}}(q^{\rho};t,q)\exp\left (\sum_{d=1}^{\infty}\frac{1}{d}\frac{Q_{R}^{d}}{q^{d/2}-q^{-d/2}}\mbox{tr}_{\tableau{1}}V^{-d} \right ).
\end{align}


The right disc contribution appears to be due to the  insertion of a $q$-brane.  This leads to a discrepancy between the disc amplitudes coming from different sides of the branes. c.f. Eq. \ref{leftdisc}. Note that our computation is valid for a stack of arbitrary number of branes, and we might as well choose a single brane.  In this case, the brane extends only on one ${\mathbb R}^{2}\subset{\mathbb R}^{4}$ in spacetime, not both. Hence, the free energy should depend only on either $q$ or $t$, not both.

We can avoid such inconsistencies if we allow holonomies to be expressed in terms of other symmetric functions with two parameter deformations.  The refined topological vertex is written in a particular basis of symmetric functions \cite{Aganagic:2012hs}, which restricts the set of possibilities. In addition to Macdonald polynomials,  we can use the dual Macdonald polynomial $Q_{\mu}(x;q,t)$ as the holonomy without spoiling the integrality of the free energy\footnote{This argument may look very mathematical without a strong physical principle behind, at least for the resolved conifold. In the next section, we give a more physical argument based on the refined Chern-Simons theory.}.  The correct holonomy turns out to be:
\begin{align}
\boxed{\mbox{tr}_{\mu_{R}}V =(-1)^{|\mu_{R}|}Q_{\mu_{R}}(V^{-1};t,q)}
\end{align}
The $\mu_{L}$-sum can be performed explicitly using the definition of the dual skew-Macdonald polynomial,
\begin{align}
\sum_{\mu_{R}} (-v^{-1}Q_{R})^{|\mu_{R}|} {\widehat N}^{\eta_{R}}_{\mu^{t}\mu_{R}}(t,q) (-1)^{|\mu_{R}|}\,Q_{\mu_{R}}(V^{-1};t,q)= Q_{\eta_{R}/\mu^{t}}(v^{-1}Q_{R}V^{-1};t,q),
\end{align}
and the $\eta_{R}$-sum leads to the correct $t$-brane contribution from the right disc:
\begin{align}\nonumber
\sum_{\eta_{R}}P_{\eta_{R}}(q^{\rho};t,q)\,Q_{\eta_{R}/\mu^{t}}&(v^{-1}Q_{R}V^{-1};t,q)\\
&=P_{\mu^{t}}(q^{\rho};t,q)\exp\left (\sum_{d=1}^{\infty}\frac{1}{d}\frac{Q_{R}^{d}}{t^{d/2}-t^{-d/2}}\mbox{tr}_{\tableau{1}}V^{-d} \right ).
\end{align}
Finally, we can do the remaining sum over $\mu$,
\begin{align}
\sum_{\mu}(-Q)^{|\mu|} P_{\mu}(t^{\rho};q,t)P_{\mu^{t}}(q^{\rho};t,q)=\prod_{i,j=1}^{\infty}(1-Q\,q^{-\rho_{i}}t^{-\rho_{j}}),
\end{align}
which gives the closed topological string amplitude of the resolved conifold. Therefore, the free energy of an internal $t$-brane, normalized by the closed amplitude, is given by
\begin{align}
F(V)=\sum_{d=1}^{\infty}\frac{1}{d}\frac{1}{t^{d/2}-t^{-d/2}}\left( Q_{L}^{d}\,\mbox{tr}_{\tableau{1}}V^{d}+Q_{R}^{d}\,  \mbox{tr}_{\tableau{1}}V^{-d}  \right )
\end{align}

The study of  integrality properties of the refined open topological string theory for even the simplest possible geometry shows that  holonomies need to chosen with some care. The integrality put such severe contraints that we can use it to determine correct holonomies. We show that for internal branes along the preferred direction, the holonomies are not the same type of Macdonald polynomials, and need to be dual with respect to the $(q,t)$-inner product of Macdonald. We will later discuss this point in more detail.

%

\subsubsection{Topological $q$-branes}
\label{sec:TopqBran}

In this subsection, we determine the required refinement of  holonomies for internal $q$-branes.  We will be very brief  since the computation does not differ much from the one of $t$-branes in the previous subsection. Eq. \ref{refinedt} is a general expression assuming the minimal change in the algebraic structure of the internal brane computation from the unrefined case and is valid for either type of branes. Let us again first focus on the ``left'' sums. We can again borrow our earlier result for the refinement of the external brane and choose the holonomy to be:
\begin{align}
\boxed{\mbox{tr}_{\mu_{L}}V=(-1)^{|\mu_{L}|}\, Q_{\mu_{L}}(V;q,t)}
\end{align}
After performing all sums for this side of the internal brane, we get
\begin{align}\nonumber
\sum_{\mu_{L},\eta_{L}} (-vQ_{L})^{|\mu_{L}|} {\widehat N}^{\eta_{L}}_{\mu\mu_{L}}&(q,t) P_{\eta_{L}}(t^{\rho};q,t)\,\mbox{tr}_{\mu_{L}}V\\
&=P_{\mu}(t^{\rho};q,t)\exp\left (\sum_{d=1}^{\infty}\frac{1}{d}\frac{Q_{L}^{d}}{q^{d/2}-q^{-d/2}}\mbox{tr}_{\tableau{1}}V^{d} \right).
\end{align}
The holomony on the other side can be easily determined to be
\begin{align}
\boxed{\mbox{tr}_{\mu_{R}}V^{-1}=v^{|\mu_{R}|}\imath P_{\mu_{R}^{t}}(V^{-1};q,t).}
\end{align}
Similarly, all sums on the right side can be performed with this chosen holonomy to get,
\begin{align}\nonumber
\sum_{\mu_{R},\eta_{R}} (-v^{-1}Q_{R})^{|\mu_{R}|} {\widehat N}^{\eta_{R}}_{\mu^{t}\mu_{R}}&(t,q) P_{\eta_{R}}(q^{\rho};t,q)\,\mbox{tr}_{\mu_{R}}V^{-1}\\
&=P_{\mu^{t}}(q^{\rho};t,q)\exp\left (\sum_{d=1}^{\infty}\frac{1}{d}\frac{Q_{R}^{d}}{q^{d/2}-q^{-d/2}}\mbox{tr}_{\tableau{1}}V^{-d} \right).
\end{align}
The remaining $\mu$-sum gives the closed topological string amplitude on the resolved conifold.

\par{Let us also note that all these different holonomies become the usual and same holonomy in the unrefined case!}


\section{Topological Branes from refined Chern-Simons Theory}\label{sec:rCS}

\par{In the previous section, we reviewed that the holonomies of the topological branes need to be modified for refined topological string theory. Moreover, we argued that the general assumption of replacing all Schur polynomials for holonomies with Macdonald polynomials leads to inconsistent free amplitudes for possible brane configurations. In this section, we adopt a different approach and derive the same conclusions from refined Chern-Simons theory.  The unrefined topological string amplitudes on local toric Calabi-Yau threefolds  were computed using Chern-Simons theory \cite{Aganagic:2000gs,Aganagic:2002qg} which shortly led to the formulation of the topological vertex \cite{Iqbal:2002we,Aganagic:2003db}. More recently, the refined Chern-Simons theory is constructed \cite{Aganagic:2011sg} and is used to compute the refined topological string amplitudes \cite{Aganagic:2012hs}. In addition to the refined $S$ and $T$ matrices, the refinement of annuli amplitudes are proposed.   }

\par{The refined Chern-Simons theory is defined as M-theory index in the following background: }
\begin{align}
(T^{*}M\times TN\times S^{1})_{q,t},
\end{align}
where the subscripts denote $q$ and $t$ denote the non-trivial fibration of Taub-NUT  space $TN$ over the M-theory circle $S^{1}$, c.f. Eq. \ref{omega}. Generically, the supersymmetry is broken, but the contangent bundle of $M$ is non-compact and there is an additional $U(1)_{R}\subset SU(2)_{R}$ that we can use to twist to preserve supersymmetry. If we wrap M5 branes in this background on
\begin{align}
(M\times \mathbb{C}\times S^{1})_{q,t},
\end{align}
we have a choice to make whether we let the M5 branes extend along the $z_{1}$ or $z_{2}$ planes! Therefore, there two distinct refinement of ordinary Chern-Simons theory.

Wilson loops along knots embedded in $M$ are non-local operators of Chern-Simons theory. We compute their contributions by considering the open topological string amplitudes. We can pick the co-normal bundle of the knot that is a Lagrangian sub-manifold $L_{K}$ in cotangent space $T^{*}M$. Additional M5 branes can be wrapped on $L_{K}$, and the theory gets a new sector from M2 branes extend from $M$ to $L_{K}$. According to this construction, $L_{K}$ has one real dimensional modulus that allows us to lift it off the base manifold $M$. Their neighbourhood looks like two Lagrangians each wrapping an $S^{1}$ in ${\mathbb C}^{*}$. M2 branes wrap the annulus of length $\Lambda$ connecting these two $S^{1}$'s on each stack of M5 branes. Their contribution to the index is captured by the annulus amplitude ${\mathcal O}(\Lambda;U,V)$ where $U$ and $V$ are the holonomies on $M$ and $L_{K}$ \cite{Ooguri:1999bv}. For the unrefined topological strings, the annuli partition function is given by
\begin{align}
{\mathcal O}(\Lambda;U,V)=\det (1-U\otimes V^{-1})^{\pm 1}.
\end{align}
\par{The exponent depends on whether the ground states of strings stretching between the D-branes are fermionic or bosonic. In topological string theory, we can distinguish between branes and anti-branes \cite{Vafa:2001qf,Aganagic:2003db}. Originally in \cite{Ooguri:1999bv}, the annulus stretches between two stacks of branes that are wrapped on three cycles which may intersect at most on an $S^{1}$. This configuration gives rise to bosonic ground states, and the associated amplitude has a $(-1)$ in the exponent. On the other hand, if one of the stacks consists of anti-branes, the ground states turn out to be fermionic; reflected in $(+1)$ in the exponent. This happens when the world volumes of branes are parallel.     }

\par{In the refined Chern-Simons theory, we can further distinguish between $q$- and $t$-branes. The refinement of  ${\mathcal O}(\Lambda;U,V)$ distinguishes all four possibilities. The annuli amplitudes due to two stacks of branes are given by }
\begin{align}
{\cal O}_{qq}(\Lambda; U,V)=\prod_{n=0}^{\infty}\frac{\det(1-q^{n}t\Lambda\, U\otimes V^{-1})}{\det(1-q^{n}\Lambda\, U\otimes V^{-1})},
\end{align}
when $q$-branes are wrapped on both $S^{3}$'s, and take the form

\begin{align}
{\cal O}_{qt}(\Lambda; U,V)=\det(1-v^{-1}\Lambda\,U\otimes V^{-1})^{-1},
\end{align}
when the branes wrapping different $S^{3}$'s are not of the same type. The annuli amplitudes due to brane and anti-brane configurations are given by
\begin{align}
{\cal O}_{q\bar{q}}(\Lambda; U,V)=\prod_{n=0}^{\infty}\frac{\det(1-q^{n}\Lambda\, U\otimes V^{-1})}{\det(1-q^{n}t\Lambda\, U\otimes V^{-1})},
\end{align}
when both branes and anti-branes are of $q$-type as the labelling suggests, and
\begin{align}
{\cal O}_{q\bar{t}}(\Lambda; U,V)=\det(1-v^{-1}\Lambda\,U\otimes V^{-1}),
\end{align}
 when one of the stacks has $q$-branes and the other $t$-branes. We refer the interested reader for the detailed derivation of the refined annuli amplitudes to \cite{Aganagic:2011sg,Aganagic:2012hs}.
\par{Let us also briefly mention that these amplitudes can be expanded in symmetric function with two parameters. For example, the annuli amplitude ${\cal O}_{q\bar{t}}(\Lambda; U,V)$ can be expanded using the Macdonald functions, }
\begin{align}
{\cal O}_{q\bar{t}}(\Lambda; U,V)=\sum_{\mu}(-v^{-1}\Lambda)^{|\mu|}P_{\mu}(U;q,t)P_{\mu^{t}}(V^{-1};t,q).
\end{align}
This expansion is by no means unique, and in the following section, we will make use of the non-uniqueness of the expansions.


\subsection{Topological $t$-branes from refined Chern-Simons theory}

\par{In the previous section, we used the integrality of the open free energy to determine correct holonomies for internal topological branes. This approach was rather mathematical and relied on known identities in the symmetric function theory. As promised, we derive  same holonomies using the refined Chern-Simons theory.  }
\par{We are studying non-compact branes, and need to specify boundary conditions at infinity to define the quantum theory. We employ by now the standard trick of considering branes wrapping on compact Lagrangian cycles. We do so without changing A-model amplitudes. To get additional $S^{3}$'s, we allow $T^{2}$ fibers degenerate over additional loci. In addition to  probe branes,   we have  other $S^{3}$'s in the geometry on which we wrap an infinite number of branes and let them go through the geometric transition. In the strict infinite number of branes limit, we take  K\"{a}hler classes of ${\mathbb P}^{1}$'s after the transition to infinite as well, leaving only a vertex.  We consider the  configuration depicted in Fig. \ref{fig:internal} to study internal $t$-branes.}
\begin{figure}[h!]
  \begin{center}
    \includegraphics[width=0.75\textwidth]{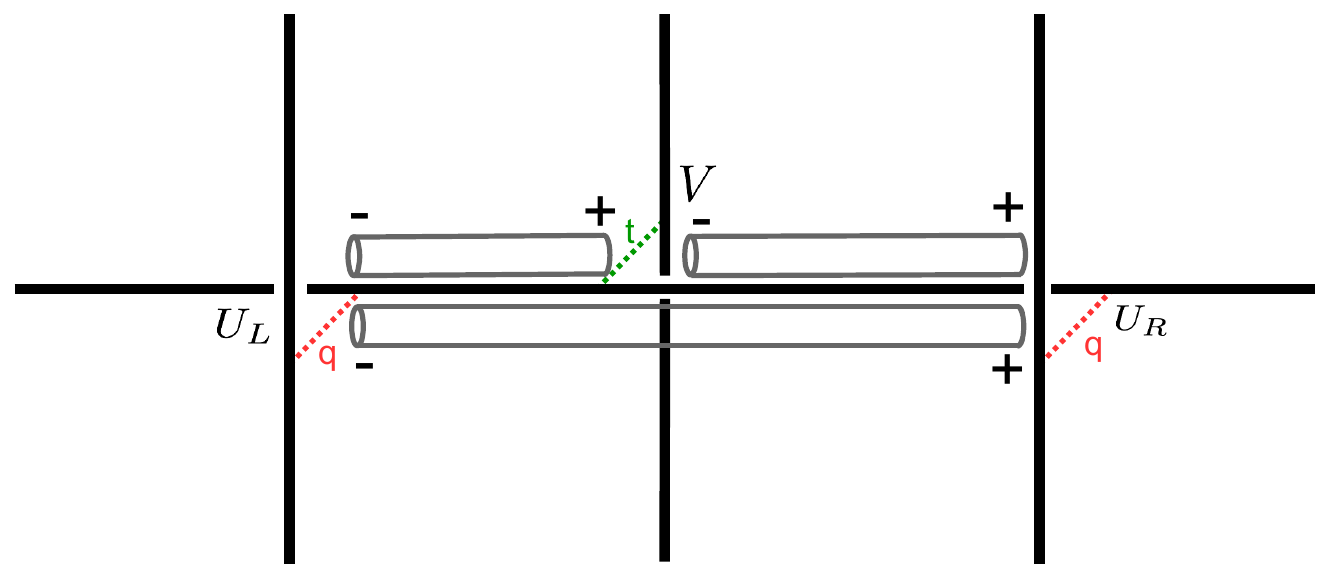}
    \caption{\label{fig:internal}The geometry before the transtion to compute the open amplitude in the presence of a stack of internal $t$-branes. }
   \end{center}
\end{figure}
\par{We wrapped $q$-branes on the left and right $S^{3}$'s and $t$-branes on the middle 3-cycle. We will let the number of $q$-branes go to infinity to get a resolved conifold with an internal brane insertion. The partition function can be obtained by computing correlators with respect to refined Chern-Simons theories on the left and right  3-cycles. Schematically, we have  }
\begin{align}
Z(V)=\langle {\cal O}_{q\bar{q}}(\Lambda;U_{R},U_{L}) {\cal O}_{qt}(\Lambda_{L};V,U_{L}) {\cal O}_{qt}(\Lambda_{R};U_{R},V) \rangle_{SU(N_{L})\otimes SU(N_{R})},
\end{align}
where we used the notation $SU(N_{L})\otimes SU(N_{R})$ to emphasize two refined Chern-Simons theories whose large $N_{L,R}$ limits we are going to take. We have already reviewed  refined annuli amplitudes from \cite{Aganagic:2011sg,Aganagic:2012hs}. These amplitudes can be expanded in various bases of symmetric functions, and we list below the ones of particular importance in our derivation,
\begin{align}
{\cal O}_{q\bar{q}}(\Lambda;U,V)&=\sum_{\mu}\Lambda^{|\mu|}\imath Q_{\mu}(U;q,t)\,P_{\mu}(V^{-1};q,t),\\
{\cal O}_{qt}(\Lambda;U,V)&=\sum_{\mu}(-v^{-1}\Lambda)^{|\mu|}\imath P_{\mu^{t}}(U;t,q)\,P_{\mu}(V^{-1};q,t)\nonumber\\
&=\sum_{\mu}(-v^{-1}\Lambda)^{|\mu|}\imath Q_{\mu^{t}}(U;q,t)\, Q_{\mu}(V^{-1};t,q).
\end{align}
Using these expansion, it is very easy to show that the partition function takes the following form,
\begin{align}\nonumber
Z(V)&=\sum_{\substack{\mu,\mu_{L},\mu_{R} \\ \eta_{L},\eta_{R}}}\Lambda^{|\mu|}(-v^{-1}\Lambda_{L})^{|\mu_{L}|}(-v^{-1}\Lambda_{R})^{|\mu_{R}|}\widehat{N}_{\mu\mu_{L}}^{\eta_{L}}(q,t)\,\langle P_{\eta_{L}}(U_{L}^{-1};q,t)\rangle_{SU(N_{L})}\\
&\times \widehat{N}_{\mu^{t}\mu_{R}}^{\eta^{t}_{R}}(t,q)\,\langle \imath Q_{\eta_{R}}(U_{R};q,t)\rangle_{SU(N_{R})}\,\imath P_{\mu^{t}_{L}}(V;t,q)\, Q_{\mu_{R}}(V^{-1};t,q).
\end{align}
To see holonomies we previously found by integrality constraints emerged from  refined Chern-Simons annuli amplitudes, we need to compute expectation values, but they have been already computed,
\begin{align}
\langle P_{\eta_{L}}(U_{L}^{-1};q,t)\rangle_{SU(N_{L})}&=P_{\eta_{L}}(t^{\rho};q,t)= (-v)^{-|\eta_{L}|}C_{\emptyset\emptyset\eta_{L}}(t,q),\\
\langle \imath Q_{\eta_{R}}(U_{R};q,t)\rangle_{SU(N_{R})}&=(-v)^{-|\eta_{R}|}P_{\eta^{t}_{R}}(q^{\rho};t,q)=C_{\emptyset\emptyset\eta^{t}_{R}}(q,t).
\end{align}
Putting everything together reproduces our results for internal $\bar{t}$ topological brane,
\begin{align}\label{expvalue}\nonumber
Z(V)&=\sum_{\mu,\mu_{L},\mu_{R}}(-v^{-1}\Lambda)^{|\mu|}(v^{-1}\Lambda_{L})^{|\mu_{L}|}(v^{-1}\Lambda_{R})^{|\mu_{R}|}C_{\emptyset\emptyset(\mu\otimes\mu_{L})}(t,q)C_{\emptyset\emptyset(\mu^{t}\otimes\mu_{R})}(q,t)\\
&\times v^{-|\mu_{L}|}\imath  P_{\mu^{t}_{L}}(V;t,q)\,(-1)^{|\mu_{R}|}Q_{\mu_{R}}(V^{-1};t,q),
\end{align}
which is the same as we proposed before if we identify
\begin{align}
Q_{L}=v^{-1}\Lambda_{L},\qquad Q_{R}=v^{-1}\Lambda_{R},\qquad Q=v^{-1}\Lambda.
\end{align}
Let us also comment on this identification. Assume we do not insert any internal branes. After the geometric transition, and letting the $N_{L,R}$ go to infinity, we end up only with the resolved conifold. For the partition function, we have
\begin{align}\nonumber
\langle {\cal O}_{q\bar{t}}(\Lambda;U,V)\rangle_{SU(N_{L,R})}&=\sum_{\mu}(-v^{-1}\Lambda)^{|\mu|}P_{\mu^{t}}(q^{-\rho};t,q)P_{\mu}(t^{-\rho};q,t)\\
&=\prod_{i,j=1}^{\infty}\left (1-Q\,q^{-\rho_{i}}t^{-\rho_{i}}\right )
\end{align}
where the first line is from the refined Chern-Simons computation, and the second from the vertex. It is clear that we need $Q\equiv v^{-1}\Lambda$!

\subsection{Topological $q$-branes from refined Chern-Simons theory}
\par{The derivation of holonomies for $q$-branes from the refined Chern-Simons theory is analogous to the one for the $t$-branes, hence we will be brief. Open topological string amplitudes can be obtained by taking the following expectation value:}
\begin{align}
Z(V)=\langle {\cal O}_{q{\bar q}} (\Lambda;U_{R},U_{L}){\cal O}_{qq} (\Lambda_{L};V,U_{L}) {\cal O}_{qq} (\Lambda_{R};U_{R},V)\rangle_{SU(N_{L})\otimes SU(N_{R})}.
\end{align}
 We will use the following expansions for $ {\cal O}_{q{\bar q}}$ and $ {\cal O}_{qq}$ in terms of Macdonald functions,
\begin{align}
 {\cal O}_{q{\bar q}}(\Lambda;U,V)&=\sum_{\mu}\Lambda^{|\mu|}\imath Q_{\mu}(U;q,t)P_{\mu}(V^{-1};q,t),\\
{\cal O}_{qq}(\Lambda;U,V)&=\sum_{\mu}\Lambda^{|\mu|}Q_{\mu}(U;q,t)P_{\mu}(V^{-1};q,t)\\
&=\sum_{\mu}\Lambda^{|\mu|}\imath Q_{\mu^{t}}(U;q,t) \imath P_{\mu^{t}}(V^{-1};q,t).
\end{align}
We can again use the expansion in symmetric functions to get,
\begin{align}\nonumber
Z(V)&=\sum_{\substack{\mu,\mu_{L},\mu_{R} \\ \eta_{L},\eta_{R}}}\Lambda^{|\mu|}\Lambda_{L}^{|\mu_{L}|}\Lambda_{R}^{|\mu_{R}|}\widehat{N}_{\mu\mu_{L}}^{\eta_{L}}(q,t)\langle P_{\eta_{L}}(U_{L}^{-1};q,t)\rangle_{SU(N_{L})}\widehat{N}_{\mu^{t}\mu_{R}}^{\eta_{R}^{t}}(t,q)\\&\times \langle \imath Q_{\eta_{R}}(U_{R};q,t)\rangle_{SU(N_{R})}\, Q_{\mu_{L}}(V;q,t) \imath P_{\mu_{R}^{t}}(V^{-1};q,t).
\end{align}
We can use the expectation value in Eq. \ref{expvalue}, and write the partition function in terms of the refined topological vertices:
\begin{align}\nonumber
Z(V)&=\sum_{\mu,\mu_{L},\mu_{R}}(-v^{-1}\Lambda)^{|\mu|}(v^{-1}\Lambda_{L})^{|\mu_{L}|}(v^{-1}\Lambda_{R})^{|\mu_{R}|}C_{\emptyset\emptyset(\mu\otimes\mu_{L})}(t,q)C_{\emptyset\emptyset(\mu^{t}\otimes\mu_{R})}(q,t)\\
&\times (-1)^{|\mu_{L}|} Q_{\mu_{L}}(V;q,t)\, v^{|\mu_{R}|} \imath P_{\mu_{R}^{t}}(V^{-1};q,t),
\end{align}
agreeing with holonomies we proposed by requiring integrality to the free energy expansion.

\subsection{Duality relation and brane changing operator}\label{dualitysection}

\par{We would like to point out a nice relation between holonomies associated to the $t$- and $q$-branes. Let us first summarize holonomies we found for internal branes:}
\begin{center}
\setlength{\extrarowheight}{4.5pt}
\begin{tabular}{c|c|c}
&left holonomy&right holonomy\\\hline
$t$-brane&$v^{-|\mu_{L}|}\imath P_{\mu^{t}_{L}}(V;t,q)$&$(-1)^{|\mu_{R}|}\, Q_{\mu_{R}}(V^{-1};t,q)$\\[1ex] \hline
$q$-brane&$(-1)^{|\mu_{L}|}\, Q_{\mu_{L}}(V;q,t) $&$v^{|\mu_{R}|} \imath P_{\mu_{R}^{t}}(V^{-1};q,t)$
\end{tabular}
\end{center}

\par{One obvious relationship is a diagonal one: the left holonomy of the $t$-brane becomes the right holonomy of the $q$-brane after a $q\leftrightarrow t$ exchange.  The same is true for remaining holonomies. }
\par{There is yet a second relationship between holonomies of the $q$- and $t$-branes that can be understood algebraically. The dual Macdonald function $Q_{\mu}(V;q,t)$ is defined using the endomorphism $\omega_{q,t}$ on the ring of symmetric functions. This is just a two parameter generalization of the involution $\omega$ on Schur functions; $\omega s_{\lambda}=s_{\lambda^{t}}$. The action of $\omega_{q,t}$ on the powersums for $t\neq 1$ is given by,}
\begin{align}
\omega_{q,t}\,p_{n}=(-1)^{n-1}\frac{1-q^{n}}{1-t^{n}}p_{n}.
\end{align}
 If we compare two holonomies, it is not hard to find a vertical relationship. Let us define the \textit{ brane changing operator} $\Omega_{t,q}$:
\begin{align}
\Omega_{t,q}\equiv (-v)^{L_{0}}\imath\, \omega_{t,q},
\end{align}
where $L_{0}$ counts the number of boxes labeling the symmetric function. The action of $\Omega_{t,q}$ can be easily computed on $t$-brane holonomies:
\begin{align}\nonumber
&\Omega_{t,q}\,\Big( v^{-|\mu_{L}|}\imath P_{\mu^{t}_{L}}(V;t,q)\Big )=\Big ( (-1)^{|\mu_{L}|}\, Q_{\mu_{L}}(V;q,t) \Big ),\\
&\Omega_{t,q}\,\Big( (-1)^{|\mu_{R}|}\, Q_{\mu_{R}}(V^{-1};t,q) \Big )=\Big( v^{|\mu_{R}|}\imath P_{\mu_{R}^{t}}(V^{-1};q,t)\Big),
\end{align}
which change the holonomies for $t$-branes into $q$-branes on each side of the brane. Hence, we reach
\begin{align}
\Omega_{t,q}:\qquad\mbox{$t$-brane}\mapsto\mbox{$q$-brane}
\end{align}

The action of  $\Omega_{t,q}$ on powersums which are isomorphic to oscillator modes is easy to determine as well:

\begin{align}
\Omega_{t,q}\,p_{n}&=\frac{t^{n/2}-t^{-n/2}}{q^{n/2}-q^{-n/2}}p_{n}
\end{align}
Note that the brane changing operator from $q$-brane to $t$-brane is just the inverse of the former operator, $\Omega_{q,t}=\Omega^{-1}_{t,q}$, similar to $\omega_{q,t}$. In addition, it become the identity operator when we take the unrefined limit, $q=t$.

\par{We can see that $\Omega_{q,t}$ is in fact the brane changing operator from refined Chern-Simons theory as well. Its action on the anti-brane is the same. The amplitude of annuli stretching between $q$ and $\bar{t}$ branes is given by,}
\begin{align}\nonumber
{\cal O}_{q\bar{t}}(\Lambda;U,V)&=\det \left (1-v^{-1}\Lambda\,U \otimes V^{-1} \right )\\
&=\sum_{\mu}(-v^{-1}\Lambda)^{|\mu|}\,P_{\mu}(U;q,t) \,P_{\mu^{t}}(V^{-1};t,q).
\end{align}
\par{We should consider the action of $\Omega_{t,q}$ on the stack of probe branes described by $P_{\mu^{t}}(V^{-1};t,q)$,  }
\begin{align}
\Omega_{t,q}\,P_{\mu^{t}}(V^{-1};t,q)=(-v)^{|\mu|}\imath Q_{\mu}(V^{-1};q,t),
\end{align}
converting the ${\cal O}_{q\bar{t}}(\Lambda;U,V)$ annulus into  ${\cal O}_{q\bar{q}}(\Lambda;U,V)$:
\begin{align}
\boxed{\Omega_{t,q}\,{\cal O}_{q\bar{t}}(\Lambda;U,V)={\cal O}_{q\bar{q}}(\Lambda;U,V)}\\
\boxed{\Omega_{t,q}\,{\cal O}_{qt}(\Lambda;U,V)={\cal O}_{qq}(\Lambda;U,V)}
\end{align}

Before we finish this section, let us briefly mention that $\Omega_{t,q}$ acts the same way on the holonomies along the un-preferred direction:
\begin{align}
 \Omega_{t,q}\left ( (-v)^{-|\lambda|}s_{\lambda}(V)\right )=\left( (-1)^{|\lambda|}S_{\lambda}(V;q,t)\right),
\end{align}
changing a $t$-brane to a $q$-brane.

\section{Application: Double compactified toric Calabi-Yau 3-fold}\label{sec:elliptic}
\par{As a principal illustration, we apply our formalism to compute `refined open topological amplitudes on the doubly compactified toric Calabi-Yau threefold with K\"{a}hler parameters $Q_{\rho}$ and $Q_{\tau} = \Lambda x Q_{m}^2$ associated with elliptic curves \cite{Hollowood:2003cv}. We study different limits of the K\"{a}hler classes with branes along the horizontal leg of the toric diagram, as depicted in Fig. \ref{fig:M-strings}.}

\begin{figure}[h!]
  \begin{center}
    \includegraphics[width=0.5\textwidth]{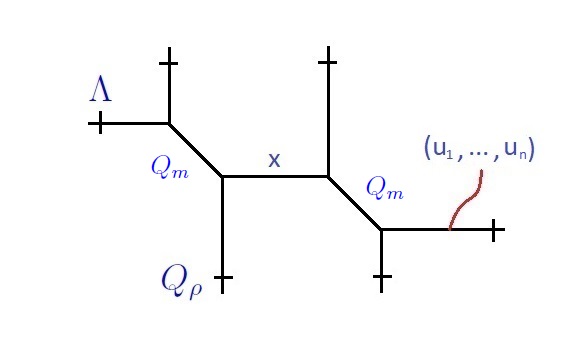}
    \caption{\label{fig:M-strings}The toric diagram assosicated with the M-strings. In addition to the vertical compactification, we also compactified the horizontal direction.}
      \end{center}
\end{figure}


This geometry generalizes the non-compact case $Q_{\rho} = Q_{\tau} = 0$, when the partition function can be straightforwardly computed with the help of standard topological vertex methods and has a number of nice properties. One can express the partition function either as a series (in what follows we fix a single brane, $n = 1$)
\begin{align}\nonumber
\frac{Z}{Z_0}\Big|_{Q_{\rho}=Q_{\tau}=0} &= 1 + \frac{\sqrt{t}(1-T)}{T(1-q)} \underbrace{\big( 1 + x Q_m \big)}_{P_1(x)} u \\
 &+ \frac{t (1 - T)(1 - q T)}{T^2 (1 - q)(1 - q^2)} \underbrace{\big( 1 + \frac{(1+q)(1-T)}{(1-qT)} x Q_m + x^2 Q_m^2 \big)}_{P_2(x)} u^2 + \ldots
\end{align}
\smallskip\\
where $T = \sqrt{\frac{q}{t}} Q_m^{-1}$, or in closed form

\begin{align}
\dfrac{Z}{Z_0}\Big|_{Q_{\rho}=Q_{\tau}=0} = \dfrac{(t^{\frac{1}{2}} u;q)_{\infty} (t^{\frac{1}{2}} u x Q_m;q)_{\infty}}{(t^{\frac{1}{2}} u / T;q)_{\infty} (t^{\frac{1}{2}} u x Q_m / T;q)_{\infty}}
\end{align}
\smallskip\\
where $(x;q)_{\infty} = \prod_{n = 0}^{\infty} (1 - q^n x)$. Both representations are interesting. In the series representation, the coefficients $P_k(x)$ are the celebrated polynomial eigenfunctions of the trigonometric Ruijsenaars-Schneider system -- Macdonald polynomials with parameters $q$ and $T$. In the closed form representation, each of the 4 factors corresponds to a BPS state in the geometry.

In this section, we will give elliptic generalizations of both representations. As it turns out, the first (series) representation generalizes most nicely to the case $Q_{\rho} \neq 0$, while the second (closed form, BPS) representation admits a nice generalization to the case $Q_{\tau} \neq 0$. This is an illustration of validity of the the holonomy prescription that we suggest.

In this geometry the edge with the brane has no other parallel edges, hence our local prescription applies and gives for the partition function of the brane

\begin{align}\nonumber
&Z =  \sum_{\{\tau_{i}\},\{\lambda_{i}\}}\sum_{\nu}  \sum_{\mu,\mu_{L},\mu_{R}} (-x)^{|\nu|}(-Q_{m})^{|\lambda_{1}|+|\lambda_{2}|}(-Q_{\tau}Q^{-1}_{m})^{|\tau_{1}|+|\tau_{2}|}(-\Lambda)^{|\mu|} \Lambda^{|\mu_{R}|}C_{\lambda_{1}^{t}\tau_{1}^{t}\nu^{t}}(q,t) \\\nonumber
&\times C_{\lambda_{1}\tau_{1}(\mu\otimes\mu_{L})}(t,q) C_{\lambda_{2}^{t}\tau_{2}^{t}(\mu^{t}\otimes\mu_{R})}(q,t)C_{\lambda_{2}\tau_{2}\nu}(t,q)\, (-1)^{|\mu_{L}|}  Q_{\mu_{L}}\big(u_1, \ldots, u_n;q,t\big) \imath P_{\mu_{R}^{t}}\big(u_1^{-1}, \ldots, u_n^{-1};q,t\big)  \\\nonumber
&=\sum_{\nu}  \sum_{\mu,\mu_{L},\mu_{R}} \sum_{\eta_{L},\eta_{R}}  (-x)^{|\nu|}(-\Lambda)^{|\mu|} \Lambda^{|\mu_{R}|}(-1)^{|\mu_{L}|}\,{\widehat N}_{\mu  \mu_{L}}^{\eta_{L}}(q,t)  {\widehat N}_{\mu^{t}\mu_{R}}^{\eta_{R}}(t,q)D_{\nu \eta_{L}}(Q_{\tau},Q_{m})\\
&\qquad\qquad\qquad\,\,\,\,\,\times D_{\eta_{R}^{t} \nu}(Q_{\tau},Q_{m}) \,  Q_{\mu_{L}}\big(u_1, \ldots, u_n;q,t\big) \imath P_{\mu_{R}^{t}}\big(u_1^{-1}, \ldots, u_n^{-1};q,t\big).
\label{Ansatz}
\end{align}
Some of the K\"{a}hler classes in the above expression are rescaled for notational purposes to make the comparison to elliptic Macdonal polynomial easier. The $\tau_{i}$ and $\lambda_{i}$ sums can be performed explicitly and we called them $D_{\nu\mu}(Q_{\tau},Q_{m})$ following \cite{Haghighat:2013gba} where it was first computed and regarded as a building block for M-strings. It takes the following form,

\begin{align}\nonumber
D_{\nu\mu}(Q_{\tau},Q_{m})&\equiv \sum_{\lambda,\tau}(-Q_{m})^{|\lambda|}(-Q_{\tau}Q_{m}^{-1})^{|\tau|}  C_{\lambda^{t}\tau^{t}\nu^{t}}(q,t)C_{\lambda\tau\mu}(t,q) \\\nonumber
&=q^{\frac{\Arrowvert\mu \Arrowvert^{2}}{2}-\frac{\Arrowvert \nu\Arrowvert^{2}}{2}}t^{-\frac{\Arrowvert \mu^{t}\Arrowvert^{2}}{2}+\frac{\Arrowvert \nu^{t}\Arrowvert^{2}}{2}}P_{\mu}(t^{-\rho};q,t)P_{\nu^{t}}(q^{-\rho};t,q)\\\nonumber
&\times\prod_{k=1}^{\infty}\prod_{(i,j)\in\nu}\frac{(1-Q_{\tau}^{k}Q_{m}^{-1}\,q^{-\nu_{i}+j-\frac{1}{2}}t^{-\mu_{j}^{t}+i-\frac{1}{2}})(1-Q_{\tau}^{k-1}Q_{m}\,q^{\nu_{i}-j+\frac{1}{2}}t^{\mu_{j}^{t}-i+\frac{1}{2}})}{(1-Q_{\tau}^{k}\,q^{-\nu_{i}+j}t^{-\nu_{j}^{t}+i-1})(1-Q_{\tau}^{k}\,q^{\nu_{i}-j+1}t^{\nu^{t}_{j}-i} )}\\
&\times\prod_{(i,j)\in\mu}\frac{(1-Q_{\tau}^{k}Q_{m}^{-1}\,q^{\mu_{i}-j+\frac{1}{2}}t^{\nu^{t}_{j}-i+\frac{1}{2}})(1-Q_{\tau}^{k-1}Q_{m}\,q^{-\mu_{i}+j-\frac{1}{2}}t^{-\nu^{t}_{j}+i-\frac{1}{2}})}{(1-Q_{\tau}^{k}\,q^{-\mu_{i}+j-1}t^{-\mu^{t}_{j}+i})(1-Q_{\tau}^{k}\,q^{\mu_{i}-j}t^{\mu_{j}^{t}-i+1})}.
\end{align}

We now proceed to compute Eq. \ref{Ansatz}. When all parameters are generic, we will compute the partition function as a series expansion in the Kahler classes in the Gopakumar-Vafa form, demonstrating that the result has correct pole structure and integrality properties. For some special values of parameters, those series expansions can be summed up in closed form and related to pronounced solutions of elliptic Ruijsenaars-Schneider integrable models: namely, the elliptic Macdonald series and dual elliptic Macdonald polynomials. The two particular cases that we consider correspond to $Q_{\tau} = 0$ (momentum-elliptic) and $Q_{\rho} = 0$ (coordinate-elliptic).

\subsection{Generic case}

As usual for any sufficiently non-trivial Calabi-Yau geometry, if all parameters are generic it is not possible to give a closed form expression for the partition function of topological strings in that background. One of the most convenient forms in which it can be expressed is the Gopakumar-Vafa form also known as the BPS expansion. We first remind how this works for a closed string partition function\footnote{If we set $\mu_{L}$ and $\mu_{R}$ to $\emptyset$ in Eq. \ref{Ansatz}; in other words, turn off the holonomies, we obtain the closed topological string partition function.}

\begin{align}
Z_0= \sum_{\mu,\nu}  (-x)^{|\nu|} \ (-\Lambda)^{|\mu|}D_{\nu \mu} (Q_{\tau},Q_{m})\, D_{\mu \nu} (Q_{\tau},Q_{m}).
\label{AnsatzClosed}
\end{align}
In this case, the Gopakumar-Vafa expansion has the form
\begin{align}
Z_0 \ = \ & \exp\left( \ \sum\limits_{d = 0}^{\infty} \dfrac{ F_0\big(q^d,t^d,x^d,\Lambda^{d},Q_m^d,Q_{\rho}^d\big) }{d(q^{d/2} - q^{-d/2})(t^{d/2} - t^{-d/2})} \ \right),
\end{align}
where the numerator $F_0(q,t,Q_{\beta})$ encodes the degeneracies, $N_{\beta}^{(j_{L},j_{R})}$, of the BPS states with their charges under the little group in 5d coming from wrapping M2 branes on $\beta$:
\begin{align}
F_0(q,t,Q_{\beta})=\sum_{\beta\in H_{2}(X,\mathbb{Z})}\sum_{j_{L},j_{R}}(-1)^{2j_{L}+2j_{R}}N_{\beta}^{(j_{L},j_{R})}\, \mbox{tr}_{j_{R}}\left (\frac{q}{t}\right )^{j_{R,3}}\, \mbox{tr}_{j_{L}}\left (q\,t\right )^{j_{L,3}}\, Q_{\beta}
\end{align}
where we used as a collective K\"{a}hler class $Q_{\beta}$ that includes $x$, $\Lambda$, $Q_{m}$, $Q_{\tau}$ and $Q_{\rho}$.
To the first few orders,
\begin{align}
\nonumber F_0 \ = \ & - x + \big( t^{\frac{1}{2}} q^{-\frac{1}{2}} + q^{\frac{1}{2}} t^{-\frac{1}{2}} \big) x Q_m - x Q_m^2 + \emph{} \\
& \nonumber \emph{} + \big( t^{\frac{1}{2}} q^{-\frac{1}{2}} + q^{\frac{1}{2}} t^{-\frac{1}{2}} \big) x Q_{\rho} Q_m^{-1}
+ \big( t^{\frac{3}{2}} q^{-\frac{3}{2}} + t^{\frac{1}{2}} q^{-\frac{1}{2}} + q^{\frac{1}{2}} t^{-\frac{1}{2}} + q^{\frac{3}{2}} t^{-\frac{3}{2}} \big) x^2 Q_{\rho} + \emph{} \\
& \emph{} - \Lambda + \big( t^{\frac{1}{2}} q^{-\frac{1}{2}} + q^{\frac{1}{2}} t^{-\frac{1}{2}} \big) \Lambda Q_m - 2 \Lambda Q_m x + \ldots.
\end{align}
Note that the result is manifestly symmetric under $q,t$ exchange, as a closed partition function should be. It is also symmetric under $(q,t) \rightarrow (q^{-1},t^{-1})$.

Similarly, for an open string partition function with a (single, $n=1$) brane insertion computed with Eq. \ref{Ansatz} and normalized over the closed part, we find,
\begin{align}
\dfrac{Z}{Z_0} \ = \ & \exp\left( \ \sum\limits_{d = 0}^{\infty} \dfrac{ F\big(q^d,t^d,u^d,x^d,\Lambda^{d},Q_m^d,Q_{\rho}^d\big) }{d(q^{d/2} - q^{-d/2})} \ \right),
\end{align}
where the numerator $F$, to the first few orders, is given by,
\begin{align}
\nonumber F \ = \ & u + u x Q_m - t^{\frac{1}{2}} q^{-\frac{1}{2}} u Q_m - t^{\frac{1}{2}} q^{-\frac{1}{2}} u x Q_m^2 - \emph{} \\
& \nonumber \emph{} - t^{-\frac{1}{2}} q^{\frac{1}{2}} u Q_m^{-1} Q_{\rho} + u x Q_m^{-1} Q_{\rho} + \big( 1 + q + t^{-1} \big) u Q_{\rho} - \emph{} \\
& \nonumber \emph{} - \big( 2 + 2 t q^{-1} + t + q^{-1} \big) t^{-\frac{1}{2}} q^{\frac{1}{2}} u Q_{\rho} x + \big( 1 + t^{-1} q + t q^{-1} \big) u Q_{\rho} x^2 + \emph{} \\
& \nonumber \emph{} + t^{\frac{1}{2}} q^{-\frac{1}{2}} \Lambda u^{-1} - \Lambda u^{-1} Q_m + t^{\frac{1}{2}} q^{-\frac{1}{2}} \Lambda x Q_m u^{-1} - x \Lambda u^{-1} Q_m^2 + \emph{} \\
& \emph{} + (1-t) q^{-\frac{1}{2}} \Lambda Q_m^2 x + \ldots.
\end{align}
Note that the result is not symmetric under $q,t$ exchange, as one can expect for an open partition function: the symmetry is broken by choosing a $q$-brane vs. a $t$-brane. This does not mean the multiplicities of BPS states depend on this choice: rather, by choosing a $q$-brane vs. a $t$-brane one counts the same BPS states with different weights. In other words, the two partition functions contain identical enumerative information.

\subsection{Case $Q_{\tau} = 0$}

In this subsection, we decompactify the ``vertical'' elliptic fiber with K\"{a}hler class $Q_{\tau}$, and have only one elliptic fiber in the ``horizontal'' direction\footnote{``Vertical'' and ``horizontal'' directions refer to the associated toric diagram.}. In the language of the integrable system literature, this choice constitutes the momentum-elliptic case.  In this case for either of the three values of $Q_m = \sqrt{\frac{q}{t}}, \sqrt{\frac{t}{q}}, \sqrt{q t}$, the following holds for the normalized open amplitude:
\begin{align}
\dfrac{Z}{Z_0} = \sum\limits_{k = 0}^{\infty} \ \left(\frac{u \sqrt{q}}{T}\right)^k \ c_k \ {\cal P}_{k}(x),
\end{align}
\smallskip\\
where $c_k$ are numeric normalization factors,
\begin{align}
c_k = \prod\limits_{i = 0}^{k-1} \prod\limits_{m = 0}^{\infty} \dfrac{(1 - Q_{\rho}^m q^{i} T)(1 - Q_{\rho}^{m+1} q^{-i} T^{-1})}{(1 - Q_{\rho}^m q^{i+1})(1 - Q_{\rho}^{m+1} q^{-i} t^{-1})},
\end{align}
and ${\cal P}_{k}(x)$ are the dual elliptic Macdonald polynomials,
\begin{align}
{\cal P}_{k}(x) = \sum\limits_{m = 0}^{k} \ \left( Q_m x\right)^m \ \prod\limits_{i = 0}^{m-1} \dfrac{\theta(q^{k-i})\theta(T q^i)}{\theta(q^{k-i}T/q)\theta(q^{i+1})},
\end{align}
with Macdonald parameters $q$ and $T = \sqrt{\frac{q}{t}} Q_m^{-1}$. Here
\begin{align}
\theta(z) = \prod\limits_{m = 0}^{\infty} (1 - Q_{\rho}^m z)(1 - Q_{\rho}^{m+1} / z)
\end{align}
is the Jacobi theta function. Note that the result only has a singularity at $q \rightarrow 1$, but not at $t \rightarrow 1$, since the choice of holonomies in Eq. \ref{Ansatz} corresponds to a (in this example single, $n=1$) $q$-brane.

Polynomial ${\cal P}_{k}(x)$ represents the brane partition function with fixed boundary condition corresponding to the $k$-th symmetric representation of $SU(2)$, and it satisfies
\begin{align}
\left( \dfrac{q Q_m x}{T} \dfrac{\theta( q^{x \partial_x - k} )}{\theta( q^{x \partial_x - k + 1} T^{-1} )} - \dfrac{\theta( q^{x \partial_x} )}{\theta( q^{x \partial_x - 1} T )} \right) \ {\cal P}_{k}(x) = 0,
\end{align}
which is the dual elliptic Ruijsenaars-Schneider equation, as given in \cite{Nieri:2015yia} Eq. 4.85 and \cite{Bullimore:2014awa} Eq. 4.59.

\subsection{Case $Q_{\rho} = 0$}

In this case, the shortest way of giving the answer is through the BPS expansion,
\begin{align}
\dfrac{Z}{Z_0} \ = \ & \exp\left( \ \sum\limits_{d = 0}^{\infty} \dfrac{ F\big(q^d,t^d,u^d,x^d,\Lambda^{d},Q_m^d,Q_{\tau}^d\big) }{d(q^{d/2} - q^{-d/2})} \ \right),
\end{align}
where the numerator $F$, to all orders, is given by,
\begin{align}
\nonumber F \ = \ \dfrac{ (1-t) q^{-\frac{1}{2}} \Lambda Q_m^2 x + (1 + x Q_m)\big( u - t^{\frac{1}{2}} q^{-\frac{1}{2}} u Q_m + t^{\frac{1}{2}} q^{-\frac{1}{2}} \Lambda u^{-1} - Q_m \Lambda u^{-1} \big)}{ 1 - Q_{\tau} },
\end{align}
where, we remind, $Q_{\tau} = \Lambda x Q_m^2$ is the total K\"{a}hler class of the compact 2-cycle. The geometric progression factor $(1 - Q_{\tau})^{-1} = 1 + Q_{\tau} + Q_{\tau}^2 + \ldots$ implies that infinitely many BPS states that wrap this cycle multiple times are organized in a simple tower of states. This is a characteristic property of the coordinate-elliptic case.

\section{Discussion and Outlook}\label{sec:discussion}

\par{In this paper, we studied the refinement of the holonomies for open topological string amplitudes. The correct holonomies are determined using the integrality properties of the refined open free energies, and are also independently derived with the refined Chern-Simons theory. Our approach demonstrates how one can place $q$- or $t$-branes on any leg of the refined topological vertex.  }

\par{The local prescription for the computation of open topological string amplitudes in the presence of internal branes is well known to be valid in the unrefined case for any toric geometry. However, in the refined case the issue is more subtle. This can be easily observed by studying surface operators in ${\cal N}=2$ $SU(2)$ theory with four flavors. In 4d, the contribution due the surface operator can be calculated by inserting an additional degenerate operator in the Liouville four-point conformal block \cite{Alday:2009fs}. If we lift the theory to 5d, $\mathbb{R}^{4}\times S^{1}$, we can study its $q$-deformation using the Dotsenko-Fateev Coulomb gas interpretation of the conformal block. From the geometric engineering view point, it is nothing but inserting an internal brane along one of the horizontal line in the toric diagram of the local $\mathbb{P}^{1}\times\mathbb{P}^{1}$ geometry blown up at four points. We used our refinement of holonomies to reproduce the surface operator contribution and observed a mismatch.}

\begin{figure}[h!]
  \begin{center}
    \includegraphics[width=0.75\textwidth]{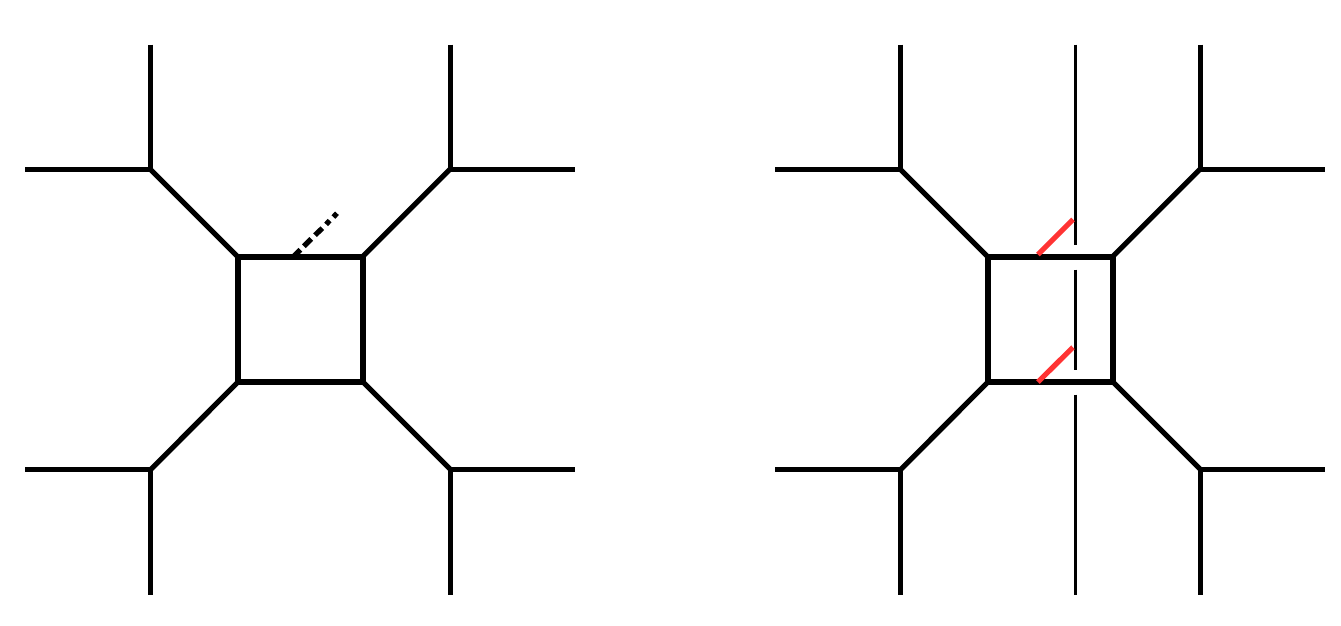}
    \caption{\label{fig:geomtran}The geometry before the transtion: we place a single brane on the upper $S^{3}$ and no branes on the lower $S^{3}$. }
   \end{center}
\end{figure}

\par{On the other hand, we used geometric transition to compute the same amplitude and found an agreement with the $q$-deformed Liouville theory computation. In this approach, we need to wrap a single brane on one of the 3-cycles for a surface operator, and none to the other 3-cycle, depicted in Fig. \ref{fig:geomtran}. The failure of the local prescription suggests that annuli can end on the 3-cycle with ``no brane'' and give contributions. This may look strange since the annuli are not expected to end on a 3-cycle if no-branes are wrapped on it. One possible explanation is that, if the same number branes and anti-branes are wrapped on it, effectively there may not be any gauge theory associated with it but the the annuli still have a place to end. This point is still under investigation and will be reported later \cite{us}. }

\par{The above observation holds true more generally. Whenever the NS5 brane used in geometric transition intersects a single $\mathbb{P}^{1}$ in the original geometry, the local prescription that we proposed reproduces the correct result. Conversely, whenever there is more than one intersection, we observe a mismatch. This suggests that further modifications may be necessary in these cases. Because of non-local nature of the ``no brane'' phenomenon described above, these modifications may ultimately lead to a non-local prescription. }

\section*{Acknowledgment}
We would like to thank Amer Iqbal and Shing-Tung Yau for useful discussions. The work of CK is supported by Department of Physics, Bo\u{g}azi\c{c}i University and CMSA, Harvard University.  CV is supported in part by NSF grant PHY-1067976.  The work of WY is supported by YMSC, Tsinghua University and CMSA, Harvard University. The work of SS is partly supported by the RFBR grant 16-02-01021.

\appendix
\section{Appenix A: Useful Identities}

\paragraph{}In this section, we collect some identities and background materials used in this note. We used the refined topological vertex to compute the open topological string amplitude which has the following explicit form,
\begin{align}
C_{\lambda\mu\nu}(t,q)=t^{-\frac{\Arrowvert\mu^{t}\Arrowvert^{2}}{2}-\frac{\Arrowvert\nu^{t}\Arrowvert^{2}}{2}}&q^{\frac{\Arrowvert\mu\Arrowvert^{2}
}{2}+\frac{\Arrowvert\nu\Arrowvert^{2}}{2}}P_{\nu}(t^{-\rho};q,t)\sum_{\eta}v^{|\eta|+|\lambda|-|\mu|}s_{\lambda^{t}/\eta}(q^{-\nu}t^{-\rho})s_{\mu/\eta}(t^{-\nu^{t}}q^{-\rho}),
\end{align}
where $P_{\nu}(t^{-\rho};q,t)$ is the Macdonald polynomial at the special point $x_{i}=t^{i-1/2}$, and $s_{\lambda/\eta}$ is the skew Schur function, defined below. Moreover, we used the shorthand notation
\begin{align}
\Arrowvert\lambda\Arrowvert^2\equiv\sum_{i=1}^{\ell(\lambda)}\lambda_{i}^{2}.
\end{align}
$v$ is given by $q^{1/2}t^{-1/2}$.
We have also made use of the following identity:
\begin{align}
P_{\nu}(t^{\rho};q,t)=(-1)^{|\nu|}q^{n(\nu^{t})}t^{-n(\nu)}P_{\nu}(t^{-\rho};q,t).
\end{align}
\par{For completeness, let us remind the reader following identities that are frequently used in our computations,}
\begin{align}
n(\lambda)&=\sum_{i=1}^{\ell(\lambda)}(i-1)\lambda_{i}=\sum_{(i,j)\in\lambda}(\lambda^{t}_{j}-i)=\sum_{(i,j)\in\lambda}(i-1)=\frac{\Arrowvert\lambda^{t} \Arrowvert^{2}}{2}-\frac{|\lambda|}{2}\\
n(\lambda^{t})&=\sum_{i=1}^{\ell(\lambda^{t})}(i-1)\lambda^{t}_{i}=\sum_{(i,j)\in\lambda^{t}}(\lambda_{i}-j)=\sum_{(i,j)\in\lambda^{t}}(j-1)=\frac{\Arrowvert\lambda\Arrowvert^{2}}{2}-\frac{|\lambda|}{2}
\end{align}
\paragraph{}In this paper, we showed that  holonomies which are describing  boundary conditions for open topological strings need to be modified in the refined case to ensure the integrality property. We argued that the naive modification from Schur functions to Macdonald functions does not capture the full story, and we need to include the dual Macdonald functions. Let us briefly recall the definition and some properties of them. Macdonald introduced a two-parameter extension of the usual inner product defined for powersum symmetric functions $p_{\mu}$, which we call $(q,t)$-inner product,
\begin{align}
\langle p_{\lambda},p_{\mu}\rangle_{q,t}=\delta_{\lambda\mu}z_{\lambda}\prod_{i=1}^{\ell(\lambda)}\frac{1-q^{\lambda_{i}}}{1-t^{\lambda_{i}}},
\end{align}
where $z_{\mu}$ is a combinatorial factor defined by
\begin{align}
z_{\lambda}=\prod_{i\geq1}i^{m_{i}}m_{i}!,
\end{align}
with $m_{i}=m_{i}(\lambda)$ being the number of rows in $\lambda$ of length $i$. $q$ and $t$ parameters introduced in refined topological strings are  same parameters as the ones introduced by Macdonald, and the unrefined case is when $q=t$, the limit when the $(q,t)$-inner product reduces to the usual inner product.

\par{}The dual Macdonald function $Q_{\mu}(x;q,t)$ is defined as the dual symmetric function of the Macdonald function with respect to the $(q,t)$-inner product:
\begin{align}
\langle P_{\lambda},Q_{\mu}\rangle_{q,t}=\delta_{\lambda\mu}.
\end{align}
There is a very nice relationship between the Macdonald polynomial $P_{\mu}(x;q,t)$ and its dual $Q_{\mu}(x;q,t)$:
\begin{align}
Q_{\mu}(x;q,t)=b_{\mu}(q,t)P_{\mu}(x;q,t),
\end{align}
where $b_{\mu}(q,t)$ is the inverse of the algebraic norm square of the Macdonald polynomial,
\begin{align}
b_{\mu}(q,t)=\frac{1}{\langle P_{\lambda},P_{\mu}\rangle_{q,t}}.
\end{align}
Another useful approach to understand the duality is obtained by generalizing the involution $\omega$ on the ring of symmetric functions similarly to depend on two parameters. Recall the action of $\omega$ on powersums $p_{\lambda}$,
\begin{align}
\omega \, p_{\lambda}=\epsilon_{\lambda}\,p_{\lambda},
\end{align}
where $\epsilon_{\lambda}=(-1)^{|\lambda|-\ell(\lambda)}$. The action of $\omega$ on Schur function is particularly interesting,
\begin{align}
\omega\,s_{\lambda}=s_{\lambda^{t}}.
\end{align}
The action of the two parameter version is defined compatible with the inner product,
\begin{align}
\omega_{q,t}\,p_{\lambda}\equiv \epsilon_{\lambda}\,p_{\lambda}\prod_{i=1}^{\ell(\lambda)}\frac{1-q^{\lambda_{i}}}{1-t^{\lambda_{i}}}.
\end{align}
The two parameter generalization of the involution $\omega_{q,t}$ relates $P_{\mu}$ and $Q_{\mu}$ as
\begin{align}\nonumber
\omega_{q,t}\,P_{\mu}(x;q,t)&=Q_{\mu^{t}}(x;t,q),\\
\omega_{q,t}\,Q_{\mu}(x;q,t)&=P_{\mu^{t}}(x;t,q).
\end{align}
The above identities follow from the simple identity $\omega_{t,q}=\omega_{q,t}^{-1}$.

On general grounds, we can easily show that the sum over all Young diagrams takes the following product form,
\begin{align}
&\sum_{\lambda}P_{\lambda}(x;q,t)Q_{\lambda}(y;q,t)=\Pi(x,y;q,t),
\end{align}
where we have,
\begin{align}
\Pi(x,y;q,t)=\omega_{t,q}\, \prod_{i,j}(1+x_{i}y_{j})=\exp\left( \sum_{n=1}^{\infty}\frac{1}{n}\frac{1-t^{n}}{1-q^{n}}p_{n}(x)p_{n}(y)\right).
\end{align}

In \cite{Awata:2011ce}, the dual Schur function $S_{\lambda}(x;q,t)$ is defined using the $(q,t)$-inner product
\begin{align}
\langle s_{\lambda},S_{\mu}\rangle_{q,t}=\delta_{\lambda\mu}.
\end{align}
The dual Schur function can also be obtained similar to the case for the Macdonald function with the help of the refined involution $\omega_{t,q}$,
\begin{align}\label{dualschur}
S_{\lambda}(x;q,t)=\imath \omega_{t,q}\,s_{\lambda}(-x),
\end{align}
and remember that $\imath$ is another involution defined by it action on the powersum symmetric function $\imath p_{\lambda}=-p_{\lambda}$.
The Cauchy identity for the Schur and dual Schur function is (sum up to the same product as the Macdonald and the dual Macdonal function),
\begin{align}
\sum_{\mu}s_{\mu}(x)S_{\mu}(y;q,t)=\Pi(x,y;q,t).
\end{align}
\paragraph{}Using Littlewood-Richardson coefficients determined by $s_{\mu}s_{\nu}=\sum_{\lambda}N_{\mu\nu}^{\lambda}s_{\lambda}$, the skew (dual) Schur function are defined by,
\begin{align}
s_{\mu/\nu}(x)=\sum_{\lambda}N_{\nu\lambda}^{\mu}s_{\lambda}(x),\\
S_{\mu/\nu}(x;q,t)=\sum_{\lambda}N_{\nu\lambda}^{\mu}S_{\lambda}(x;q,t).
\end{align}
The dual Schur functions satisfy similar to the Schur functions,
\begin{align}
S_{\mu}(x,y;q,t)=\sum_{\lambda}S_{\mu/\lambda}(x;q,t)S_{\lambda}(y;q,t).
\end{align}

The skew (dual) Macdonald function are defined after refining the Littlewood-Richardson coefficients which we denote by ${\widehat N}_{\nu\eta}^{\mu}(q,t)$. There are defined similar to Schur function using the Macdonald polynomials,
\begin{align}
P_{\nu}(x;q,t)P_{\eta}(x;q,t)=\sum_{\mu}{\widehat N}_{\nu\eta}^{\mu}(q,t)P_{\mu}(x;q,t)
\end{align}
We list the identities required for our computations without any proofs,
\begin{align}
Q_{\mu/\nu}(x;q,t)&=\sum_{\eta}{\widehat N}_{\nu\eta}^{\mu}(q,t)Q_{\eta}(x;q,t),\\
P_{\mu/\nu}(x;q,t)&=\sum_{\eta}{\widehat N}_{\nu^{t}\eta^{t}}^{\mu^{t}}(t,q)P_{\eta}(x;q,t).
\end{align}
The Cauchy identities are deformed a little bit once we include the skew functions,
\begin{align}
&\sum_{\lambda}P_{\lambda}(x;q,t)Q_{\lambda/\mu}(y;q,t)=P_{\mu}(x;q,t)\Pi(x,y;q,t),\\
&\sum_{\lambda}P_{\lambda^{t}/\mu}(x;t,q)P_{\lambda}(y;q,t)=P_{\mu^{t}}(y;q,t)\prod_{i,j}(1+x_{i}y_{j}).
\end{align}

\bibliography{references}
\end{document}